\documentclass[conference]{IEEEtran}

\IEEEoverridecommandlockouts
\usepackage{amsmath,amssymb}
\usepackage{graphicx}
\usepackage{booktabs}
\usepackage{cite}

\usepackage{microtype}
\usepackage{xcolor}
\usepackage[hidelinks]{hyperref}

\newcommand{\bits}{\{0,1\}^n}

\title{Reliable Sample-Level Quantum Error Mitigation\\ via Dominance-Aware Clustering}

\author{%
\IEEEauthorblockN{Mohsen Ghodrati, Kausthubh Chandramouli, and Dror Baron}
\IEEEauthorblockA{North Carolina State University\\
Raleigh, NC, USA\\
\{mghodra, kprabha, barondror\}@ncsu.edu}
\thanks{This work was supported in part by the Institute for Robust Quantum Simulation OMA--2120757.}}

\begin{document}
\maketitle

\begin{abstract}
Many quantum algorithms for classically difficult optimization tasks must return high-quality bitstrings from finitely many circuit executions, whereas most quantum error-mitigation methods target expectation values.
We study sample-level recovery when measured probability mass is distributed around multiple latent bitstrings, called centers.
Each component of the measured probability mass is called a source and we assume that each center is associated with one source.
We identify dominance---at every coordinate, more than half of a retained region's probability mass comes from one source and agrees with its center---as a sufficient condition under which majority voting recovers that center with exponentially decreasing error probability.
We show that nearest-center assignment, as used in clustering algorithms such as the $k$-modes algorithm, can fail to produce dominated regions even when the true centers are known.
This failure motivates responsibility thresholding and a local dominance screen, whose combination we call dominance-aware (DA) refinement.
Synthetic and simulated MaxCut--QAOA experiments show that DA refinement favors precision, while $k$-modes with DA refinement improves overall center recovery.
All procedures are classical post-processing and require no additional quantum-circuit executions.
\end{abstract}

\begin{IEEEkeywords}
Center discovery, clustering, quantum error-mitigation, sample-level inference
\end{IEEEkeywords}

\section{Introduction}
Quantum algorithms are pursued for classically difficult tasks such as combinatorial optimization, but hardware noise, finite circuit depth, and finite execution budgets can prevent their measured bitstrings, called \emph{shots}, from revealing useful solutions.
Most quantum error-mitigation (QEM) methods improve expectation values, i.e., measured averages, over shots \cite{temme_error_2017}. %,Endo2018,Cai2023}. 
In algorithms such as the Quantum Approximate Optimization Algorithm (QAOA), however, the measured bitstrings are themselves the object of interest \cite{farhi_quantum_2014}; correcting an average does not identify which solutions should be returned. 
Sample-level QEM therefore seeks task-relevant bitstrings from a finite measured distribution.

For a computational task, we call the task-relevant bitstrings $c_1,\ldots,c_K\in\bits$ \emph{centers}.
In MaxCut, for example, where the task is to partition the vertices of a graph into two groups to maximize the number of edges crossing between them, the centers are the bitstrings encoding globally optimal cuts \cite{farhi_quantum_2014}. 
A center need not be observed exactly or be the most frequent measured bitstring.
This distinction matters for methods such as matrix-free measurement mitigation (M3) \cite{nation_scalable_2021}, which restricts sample-level QEM to the subspace of bitstrings observed in the data.
To motivate the challenge of recovering unobserved centers, consider a single-source model, where each bit differs independently from the underlying center with probability $\epsilon<1/2$. 
Directly observing the center with some fixed confidence level requires a number of shots that grows exponentially with bitstring length $n$. 
In contrast, qubit-wise majority voting (QMV), which selects the majority value at each bit, can estimate the center with the same confidence using only logarithmically many shots in $n$ \cite{baron_qubit-wise_2024}. 
This exponential-versus-logarithmic gap highlights the advantage of QMV under finite-shot budgets.
The present work identifies a sufficient condition for extending this logarithmic-in-$n$ guarantee to clusters containing shots from multiple latent sources associated with different centers: a cluster is \emph{dominated} by the \emph{source} $k$, whose associated center is $c_k$, when shots from the source $k$ that agree with $c_k$ contribute more than half of the retained probability at every coordinate.

Two recent methods address sample-level QEM. One approach, Q-Cluster, assigns each shot to the candidate center with minimum \emph{Hamming distance}, the number of differing bits, and updates each center by QMV \cite{patil_q-cluster_2025}. 
Q-Cluster's geometric assignment ignores unequal mixture weights and can produce a cluster that is not dominated even when the candidate centers are correct.
The second approach, EM-QEM, is an expectation-maximization (EM)-based quantum error-mitigation method that first filters shots consistent with diffuse background noise and then alternates between soft source assignment and parameter updates to estimate centers, source weights, and bit-flip rates \cite{chandramouli_statistical_2025}.
EM-QEM accounts for unequal mixture weights, but the nonconvex likelihood landscape that EM optimizes over can produce initialization-dependent local solutions. 
Although both Q-Cluster and EM-QEM can recover unobserved centers, their analyses do not give a finite-sample condition under which a cluster containing shots from multiple sources supports exact center recovery.

\vspace{1em}
\noindent\textbf{Contributions.}
We formulate sample-level QEM as a maximum-likelihood recovery problem for task-relevant bitstrings under a global-plus-local mixture model.
Dominance is then identified as a sufficient condition under which QMV recovers a center from a cluster using a number of assigned shots that grows only logarithmically with the bitstring length, for a fixed confidence and dominance margin.
We further show that \emph{nearest-center Hamming assignment}, which assigns each bitstring to its closest candidate center in the Hamming space, can produce clusters that do not support reliable QMV recovery even when the true centers are given.
The failure of nearest-center assignment motivates thresholding fitted source probabilities, called responsibilities, and a local dominance screen. Their combination defines dominance-aware (DA) refinement.
We instantiate DA refinement in the $k$-modes algorithm and introduce Adaptive DA $k$-modes as an end-to-end sample-level QEM method.

\section{Model and MLE formulation}
For a computational task $\mathcal{T}$, e.g., MaxCut, let $C=\{c_1,\ldots,c_K\}\subseteq\bits$ be a set of $K\ge1$ distinct task-relevant bitstrings of $\mathcal{T}$, called \emph{centers}.
For a quantum algorithm $\mathcal{Q}$ that attempts to solve $\mathcal{T}$, we model the measured distribution of $\mathcal{Q}$ as a probabilistic mixture model with
\begin{equation}
P_{\theta}(x)
=
\frac{\alpha_0}{2^{n}}
+
\sum_{k=1}^{K}\!\alpha_k
\prod_{i=1}^{n}
\!\epsilon_{k,i}^{\,x_i\oplus(c_k)_i}
(1\!-\!\epsilon_{k,i})^{1-x_i\oplus(c_k)_i},
\label{eq:explicit-mixture}
\end{equation}
where $x\in\bits$, $x_i$ and $(c_k)_i$ are the $i$-th bits of $x$ and $c_k$, respectively, $\alpha_k\geq0$ for $k=0,\dots,K$, $\sum_{k=0}^{K}\alpha_k=1$, $0\leq\epsilon_{k,i}<1/2$, and $\theta=(C,\boldsymbol{\alpha},\boldsymbol{\epsilon})$.
We assume that executing $\mathcal{Q}$ produces $S$ i.i.d. shots $X^{(s)}\in\{0,1\}^n$, $s\in\{1,\ldots,S\}$, drawn from \eqref{eq:explicit-mixture}.
Let $G^{(s)}\in\{0,1,\ldots,K\}$ denote the latent \emph{source} of $X^{(s)}$:
the component of the mixture in \eqref{eq:explicit-mixture} that produced $X^{(s)}$. 
The source $0$ is the uniform background, while the source $k\ge1$ is centered on $c_k$ and generates a noisy version of it, hence
$\Pr(G^{(s)}\!=\!k)=\alpha_k$.
For $k\ge 1$, we call $c_k$ the associated center with the source $k$.
Conditioned on $G^{(s)}=k\geq1$, we assume that the $i$-th bit of $X^{(s)}$ differs from $(c_k)_i$ with probability $\epsilon_{k,i}<1/2$, independently across the bits and shots. 
We call $\epsilon_{k,i}$ the \emph{bit-flip rate} at the coordinate $i$ of the source $k$.

The uniform term in \eqref{eq:explicit-mixture} models an effective global depolarizing contribution, motivated by results showing that sufficiently scrambling noisy circuits can produce an approximately uniform output component \cite{dalzell_random_2024}.%
\footnote{The uniform term represents background probability in the measured output.%
% It does not mean that the physical hardware noise is always exactly depolarizing.
} 
The local terms model perturbations around task-relevant bitstrings \cite{baron_qubit-wise_2024,patil_q-cluster_2025,chandramouli_statistical_2025,foldager_can_2024}.

For a fixed $K$, the sample-level QEM can be formulated as a joint maximum-likelihood optimization problem,
\begin{equation}
\widehat{\theta}_K
\in
\arg\max_{\theta\in\Theta_K}
\frac{1}{S}\sum_{s=1}^{S}
\log P_{\theta}\!\left(X^{(s)}\right),
\label{eq:empirical-mle}
\end{equation}
where $\Theta_K$ is the feasible parameter set.
Equation~\eqref{eq:empirical-mle} jointly estimates the centers, the source weights, and the source-specific bit-flip rates.
A restricted two-center version of the discrete search reduces to the NP-hard hypercube \(2\)-segmentation problem \cite{feige_np-hardness_2014}, motivating our dominance-aware approximate method; this hardness claim does not apply to every individual parameter update.

% For a fixed $K$, the sample-level QEM can be formulated as a joint maximum-likelihood optimization problem,
% \vspace{-0.7em}
% \begin{equation}
% \widehat{\theta}_K
% \in
% \arg\max_{\theta\in\Theta_K}
% \frac{1}{S}\sum\nolimits_{s=1}^{S}
% \log P_{\theta}\!\left(X^{(s)}\right),
% \label{eq:empirical-mle}
% \end{equation}
% where $\Theta_K$ is the feasible parameter set.
% Equation~\eqref{eq:empirical-mle} jointly estimates the centers, the source weights, and the source-specific bit-flip rates.
% Even simplified versions of the resulting discrete center-optimization problem reduce to known NP-hard problems\cite{Feige2014Hypercube}, motivating our dominance-aware approximate method.

\section{Dominance and finite-shot recovery}
Fix a region $\mathcal A\subseteq\{0,1\}^n$ and let $\gamma_k(\mathcal A)\!:=\!\Pr(G\!=\!k\!\mid\! X\!\in\!\mathcal A)$.
Conditioning on $\mathcal A$ can change the effective error pattern of a source.
Define the conditional bit-flip rates
\begin{equation}
\epsilon_{k,i}^{\mathcal A}
:=
\Pr\!\left(
X_i\!\neq\!(c_k)_i
\mid
G\!=\!k,\ X\!\in\!\mathcal A
\right).
\label{eq:epsilon-local}
\end{equation}
Define
\begin{equation}
\Delta_{k,i}^{\mathcal A}:=\gamma_k(\mathcal A)(1-\epsilon_{k,i}^{\mathcal A})-1/2.
% = \Pr(G\!=\!k,\,X_i\!=\!(c_k)_i\!\mid\! X\!\in\!\mathcal A)-\frac{1}{2}
\label{eq:coordinate-dominance-margin}
\end{equation}
For every coordinate $i$ of a fixed center $c_k$,
\begin{equation}
\Pr\!\left(
X_i\!=\!(c_k)_i\mid X\!\in\!\mathcal A
\right)
\ge
\frac12+\Delta_{k,i}^{\mathcal A}.
\label{eq:target-source-margin}
\end{equation}
A \emph{region} $\mathcal A$ is a subset of the bitstring space, whereas $\mathcal S(\mathcal A):=\{X^{(s)}:X^{(s)}\in\mathcal A\}$ is the corresponding observed \emph{cluster}, understood as a multiset of shots.
We call the region $\mathcal A$, and equivalently its observed cluster $\mathcal S(\mathcal A)$, \emph{dominated (by the source $k$)}
% (or by its center $c_k$) 
when
% \begin{equation*}
$
\Delta_k^{\mathcal A}:=\min_i\Delta_{k,i}^{\mathcal A}>0
$.
% \end{equation*}
We call $\Delta_k^{\mathcal A}$ the \emph{dominance score} of source $k$ in region $\mathcal A$.
Let $\widehat c_k=\operatorname{QMV}[\mathcal S(\mathcal A)]$ denote the bitstring obtained by coordinate-wise majority vote in $\mathcal S(\mathcal A)$.
For $S_{\mathcal A}$ independent shots assigned to a region $\mathcal A$ dominated by source $k$, Hoeffding's inequality and the union bound give
\begin{equation}
\Pr(\widehat c_k\neq c_k)
\le
n\exp\left(-2S_{\mathcal A}{(\Delta_k^{\mathcal A})}^2\right).
\label{eq:qmv-bound}
\end{equation}
Hence, $S_{\mathcal A}\ge \tfrac12\log(n/\delta){(\Delta_k^{\mathcal A})}^{-2}$ suffices for failure probability at most $\delta$.
The dominance condition is sufficient rather than necessary because shots from other sources may also agree with $c_k$.

\vspace{1em}
\noindent\textbf{Failure of nearest-center assignment.}
Nearest-center assignment can bias a QMV update even when the candidate centers are exactly correct. 
For example, with centers $(000,001,111)$, mixture weights $(0.95,0.04,0.01)$, and a symmetric bit-flip rate $0.1$, a nearest-center assignment assigns the region $\{110,111\}$ to the center $111$, yet the majority of shots in that region are $110$.
In this example, the responsibility thresholding criterion introduced in the following retains shots equal to $111$ while rejecting those equal to $110$.

To see when this failure disappears, assume equal source weights, no background probability mass, and
\(\epsilon_{k,i}\le 1/2-\beta\) for some \(\beta>0\).
Let \(d_{\min}\) be the minimum pairwise Hamming distance between the centers.
For a shot from source \(k\), assignment to another center requires at least half of the coordinates on which the two centers differ to change.
Hoeffding's inequality bounds this event by
\(q=\exp(-2\beta^2d_{\min})\); a union bound over the competing centers bounds the total leakage by \(Kq\).
If \(q<\beta/K\), the source-\(k\) probability mass that remains in its nearest-center region and agrees with \(c_k\) exceeds one half at every coordinate.
Equivalently,
\(d_{\min}>(2\beta^2)^{-1}\log(K/\beta)\)
is sufficient.
Thus, for fixed \(\beta\), dominance is automatic once the center separation is of order \(\log K\), and the QMV guarantee in \eqref{eq:qmv-bound} applies to every nearest-center region.
This result explains why Q-Cluster can succeed when the latent centers are sufficiently well separated.

% In the longer paper, we prove that, under equal mixture weights, no diffuse background, and a fixed small symmetric bit-flip rate; pairwise Hamming distance proportional to $\log(K)$ is sufficient at the population level for nearest-center regions to support exact QMV recovery.
% This separation result explains why Q-Cluster can succeed when the centers are sufficiently well separated.

\section{Dominance-aware refinement}
Define the conditional likelihood of $X$ under source $k$ as
% \begin{equation*}
\begin{equation}
L_k(X):=\prod_{i=1}^{n}
\epsilon_{k,i}^{\,X_i\oplus(c_k)_i}
(1-\epsilon_{k,i})^{1-X_i\oplus(c_k)_i}.
\end{equation}
% \end{equation*}
The \emph{responsibility} of source $k$ for the shot $X$ is
\begin{equation}
r_k(X)\!:=\!\Pr(G\!=\!k\!\mid\! X)
\!=\!\frac{\alpha_kL_k(X)}
{{\alpha_0}{2^{-n}}+\sum_{\ell=1}^{K}\alpha_\ell L_\ell(X)}.
\label{eq:source-responsibility}
\end{equation}
For a threshold $\lambda>1/2$, let
\begin{equation}
\mathcal A_k(\lambda)
:=
\{x\in\bits:r_k(x)>\lambda\}.
\label{eq:A-k-lambda}
\end{equation}
If $\Pr\left(X\in\mathcal A_k(\lambda)\right)>0$, then
\begin{equation}
\gamma_k(\mathcal A_k(\lambda))
=
\mathbb E[
r_k(X)\mid X\in\mathcal A_k(\lambda)
]
>\lambda.
\label{eq:threshold-increases-source-fraction}
\end{equation}
Define $\epsilon_{k,i}^{\mathcal A_k(\lambda)}$ analogously to \eqref{eq:epsilon-local}. 
If
\begin{equation}
\Delta_k^{\mathcal A_k(\lambda)}\!:=\!\min_{1\le i \le n}\!
\left\{
\gamma_k\left(\mathcal A_k(\lambda)\right)
\!\left(
1\!-\!\epsilon_{k,i}^{\mathcal A_k(\lambda)}
\right)
\!-\!\frac12
\right\}
>0,
\label{eq:dominance-screening}
\end{equation}
then the QMV bound \eqref{eq:qmv-bound} applies with $\Delta_k^{\mathcal A_k(\lambda)}$.
At the population level, responsibility thresholding in \eqref{eq:threshold-increases-source-fraction} and a positive score in \eqref{eq:dominance-screening} imply the QMV bound \eqref{eq:qmv-bound}. In practice, fitted responsibilities define the regions, and the corresponding responsibility-weighted agreement score is used as an empirical screen rather than as a population certificate. The two operations can refine candidate centers or clusters; we call their combination \emph{dominance-aware (DA) refinement}.

DA refinement is a module initialized with an estimate $\hat K$ of $K$, candidate centers $\widehat C$ or candidate clusters $\{\widehat{\mathcal S}_k\}_{k=1}^{\hat K}$, and, for center inputs, thresholds $\lambda_k>1/2$, where $k=1,\ldots,\hat K$. It performs the following steps:

\noindent\emph{(i)} For candidate centers, fits the mixture weights $\widehat{\boldsymbol{\alpha}}$ and bit-flip rates $\widehat{\boldsymbol{\epsilon}}$ with $\widehat C$ fixed. It then computes the responsibilities in \eqref{eq:source-responsibility}, forms $\widehat{\mathcal A}_k$ according to \eqref{eq:A-k-lambda} using $\lambda_k$, and sets $\widehat{\mathcal S}_k:=\mathcal S(\widehat{\mathcal A}_k)$, for $k=1,\ldots,\hat K$. For candidate clusters, it retains the initialized clusters $\widehat{\mathcal S}_k$.

\noindent\emph{(ii)} Obtains $\widetilde c_k=\operatorname{QMV}[\widehat{\mathcal S}_k]$, retaining the previous center for an empty cluster and merging repetitions.

\noindent\emph{(iii)} Fits provisional parameters with $\widetilde C$ fixed and recomputes the responsibility regions and induced clusters used for screening.

\noindent\emph{(iv)} Accepts a tentative center only if its fitted empirical dominance score---the minimum responsibility-weighted coordinate agreement within its recomputed region, minus $1/2$---is positive; otherwise, it restores the corresponding center and may increase $\lambda_k$. After merging repetitions, it refits with $\widehat C^{+}$ fixed.

\vspace{1em}
\noindent\textbf{Adaptive and Lightning DA \texorpdfstring{$k$}{k}-modes}
We instantiate DA refinement in the $k$-modes algorithm
\cite{huang_extensions_1998}, because $k$-modes follows the same iterative
assign-and-update structure as Q-Cluster and EM-QEM.
For a returned center set $\widehat C$ and reference set $C^\star$, define
\begin{equation}
P_C\!:=\!\frac{|\widehat C\!\cap\! C^\star|}{|\widehat C|},
\
R_C\!:=\!\frac{|\widehat C\!\cap\! C^\star|}{|C^\star|},
\
F_1^C\!:=\!\frac{2P_CR_C}{P_C\!+\!R_C},
% \footnote{We set $F_1^C=0$ when $P_C+R_C=0$, including when $\widehat C=\varnothing$, and set $P_C=0$ when $\widehat C=\varnothing$.}
\label{eq:center-set-f1}
\end{equation}
where $P_C$ and $R_C$ are called the \emph{precision} and \emph{recall} 
% rates 
of
$\widehat C$, respectively. %
% \footnote{Precision measures how many returned centers are correct, recall measures how many reference centers are recovered, and $F_1^C$ balances the two.}
DA refinement is specifically designed to favor precision, but cannot recover a center whose recovery basin contains no initialized candidate.
% \emph{Adaptive DA $k$-modes} addresses this recall limitation by pruning an \emph{overcomplete fit}, a fit initialized with more candidates than are ultimately retained, and jointly refining diverse candidates proposed from concentrated probability mass poorly explained by the current fit.
An \emph{overcomplete fit} is a fit initialized with more candidates than are ultimately retained.
\emph{Adaptive DA $k$-modes} addresses DA refinement's recall limitation by pruning such a fit and jointly refining diverse candidates proposed from concentrated probability mass poorly explained by the current fit.
Adaptive DA $k$-modes sets its proposal and pruning parameters from the 
% observed entropy and the fraction of 
observed bitstrings, allowing more candidates when the data are diffuse.
To reduce computational cost, \emph{Lightning DA $k$-modes} retains the same candidate-proposal and DA-refinement structure but reduces its
% warm-started 
parameter updates per fit.
All processing remains classical. 
The implementation details are deferred to the full paper.

\section{Synthetic and QAOA evaluation}
We evaluate center recovery using candidates constructed only from the
observed shots. No method receives the true number of centers
(unknown-$K$), the task objective, or the reference centers, and each
method's hyperparameters are fixed across all datasets within a benchmark.

A stream is one nested shot sequence generated for a fixed setting and seed.
The small-\(n\) synthetic benchmark uses \(n=10\), \(K=4\), \(150\)
\emph{marginal} 
streams that vary one generating factor at a time, and \(243\) 
\emph{factorial} 
streams that jointly vary background probability mass, source imbalance, bit-flip probabilities, and center separation.
We report the largest tested budget, \(S=32{,}768\).

The QAOA benchmark uses \(26\) \(n=10\) MaxCut instances from six graph structures, two QAOA layers, and \(S=16{,}384\).
It includes no added hardware noise, weak output noise with \(6\%\) uniform replacement and coordinate-dependent bit-flip probabilities from \(0.02\) to \(0.07\), and an IBM-backend noise simulation.
Because QAOA candidates may be checked by the task objective, we report \(R_6\), the recall after checking at most the first six ranked candidates, averaged over the graph structures and noise conditions.

\begin{figure*}[t]
\centering
\begin{minipage}[t]{0.49\linewidth}
\centering
\includegraphics[width=0.8\linewidth]{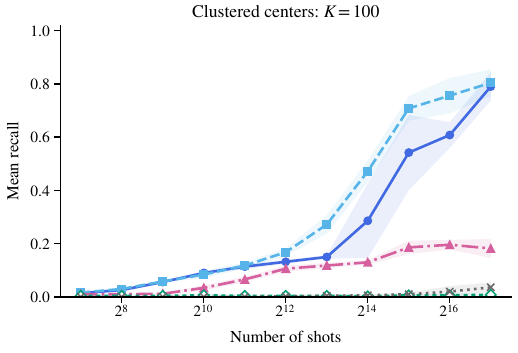}
\end{minipage}\hfill
\begin{minipage}[t]{0.49\linewidth}
\centering
\includegraphics[width=0.8\linewidth]{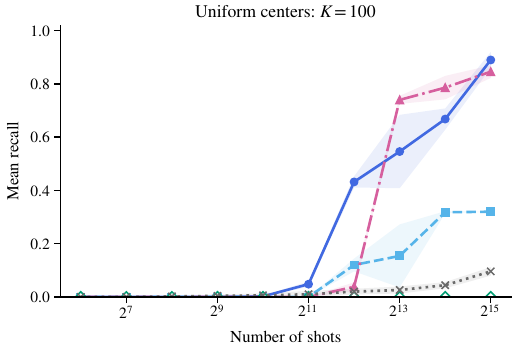}
\end{minipage}\\[-0.25em]
\includegraphics[width=\linewidth]{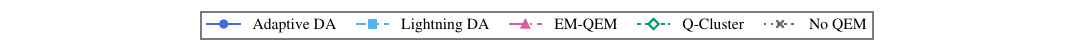}
\caption{Unknown-\(K\) center recovery for \(n=100\) and \(K=100\).
Left: clustered centers. Right: uniformly drawn centers.
Curves show mean recall and \(95\%\) confidence intervals over five independent streams.
The clustered and uniform sweeps end at \(S=131{,}072\) and \(S=32{,}768\), respectively.}
\label{fig:highdim-k100-conference}
\end{figure*}

The high-dimensional benchmark uses \(n=100\), \(K=100\), and five independent streams per geometry.
Clustered centers form ten groups of ten and use \(S=131{,}072\); uniformly drawn centers use \(S=32{,}768\).
For evaluation, \(C^\star\) contains the true synthetic centers or all retained globally optimal cuts, but no method receives \(C^\star\) or the MaxCut objective.
Fig.~\ref{fig:highdim-k100-conference} shows the complete unknown-\(K\) shot sweeps for the two \(n=100\) geometries.

At \(n=10\), Adaptive DA has the highest 
marginal 
\(F_1^C\) and remains within \(0.02\) of EM-QEM on the 
factorial 
design; Lightning DA also reaches \(F_1^C=0.72\) on the 
marginal 
design.
With six QAOA candidate checks, Adaptive DA has the highest average recall, \(0.65\).
The \(n=100\) results show a stronger geometry-dependent advantage.
For clustered centers, Lightning and Adaptive DA reach recall \(0.80\) and \(0.79\), compared with \(0.18\) for EM-QEM and \(0.01\) for Q-Cluster.
For uniform centers, Adaptive DA reaches recall \(0.89\), compared with \(0.85\) for EM-QEM, \(0.32\) for Lightning DA, and \(0.00\) for Q-Cluster.
Thus, Lightning DA is effective when local clustered structure supplies candidates in one round, whereas repeated Adaptive DA proposals are important for dispersed centers.

\section{Scope and conclusion}
Reliable center recovery depends not only on clustering nearby shots but also on whether each retained cluster supports QMV. DA refinement combines responsibility-thresholded regions with a local dominance screen to reject unsupported QMV updates.
DA refinement requires no additional quantum-circuit executions, gives a QMV-based recovery mechanism with logarithmic verification complexity, and remains meaningful in regimes where exact center observation is impossible or where naive frequency baselines fail.
At the selected low-dimensional budgets, Adaptive DA is competitive in full-set recovery and leads the verification-limited QAOA comparison; at \(n=100\), Adaptive or Lightning DA give the highest recall for both tested center geometries.

% \IEEEtriggeratref{5}
\bibliographystyle{IEEEtran}
\bibliography{IEEEabrv,references}

\end{document}